\documentclass{icbo}

\usepackage{listings}
\pdfoutput=1
\usepackage{natbib}
\usepackage{url}
\usepackage{clojure}
\usepackage{tawny}

\lstnewenvironment{code}[1][]%
{\lstset{style=tawnystyle,frame=shadowbox,#1}}{}
\lstMakeShortInline[style=tawnystyle]|

\title[Short Version of Title]{Identitas: A Better Way To Be Meaningless}
\author[Alshammry \textit{et~al}]{ Nizal Alshammry\,$^1$\,$^,$\,$^2$\footnote{To whom correspondence should be addressed: N.K.E.Alshammry2@newcastle.ac.uk} and Phillip Lord\,$^1$\,}
\address{$^{1}$School of Computing Science,  Newcastle University NE1 7RU, United Kingdom\\
$^{2}$Department of Computing Science, Northern Borders University, Saudi Arabia}

\begin{document}

\maketitle

\begin{abstract}
  It is often recommended that identifiers for ontology terms should
  be semantics-free or meaningless. In practice, ontology developers
  tend to use numeric identifiers, starting at 1 and working
  upwards. Here we describe a number of significant flaws to this
  scheme, and the alternatives to them which we have implemented in
  our library, identitas.

  Software is available from \url{https://github.com/phillord/identitas}.
\end{abstract}

During the years that ontologies have moved to becoming a standard
part of the biomedical chain, a set of standard practices have build
up which are used to enable their good management, including the
addition of standardised metadata about each ontology term, including
labels, definitions, editorial status and so forth.

One key piece of metadata is the identifier. For most ontological
technologies this is in the form of an IRI (Internationalized Resource
Identifer), or something that is convertable into one. Much has been
written about the nature of identifier and how they should be
chosen. The percieved wisdom is that identifiers should be
\emph{semantics-free} or meaningless. The key
aim here is to enable persistence of access to a
term~\citep{greycite81775}; an identifier which is based on some
semantics associated with the term may need to be changed when that
aspect changes, even if the change does not reflect a change in the
ontological semantics.

As an example, OBO Foundry principles~\cite{OBOFoundry2008} provide
guidelines for identifiers; these include both management principles
(``The ID-space / prefix must be registered with the OBO library in
advance.''), syntactic constraints (``The URI should be constructed
from a base URI, a prefix that is unique within the Foundry (e.g. GO,
CHEBI, CL) and a local identifier (e.g. 0000001).''), in addition to a
strong commitment to semantics-free IDs (``The local identifier should
not consist of labels or mnemonics meaningful to humans.''). No
specific advice is given on the form of the local identifier; however,
in practice OBO identifiers use numeric IDs, 8 numerals long,
approximately increasing monotonically.

While semantics-free identifiers have their advantages there are
distinct disadvantages as well, especially for humans. They are poorly
mnemonic, hard to differentiate from each other and relatively
difficult to read. For this reason, for example, many bioinformatics
databases provide both semantic-free \emph{accession numbers} (which
are essentially the same thing as an identifer in ontology
terminology), and an \emph{identifier} (which is rather like a
compressed, syntactically predicatable label). It is also interesting
to note that, with software development, programmers emphasise the use
importance of semantically-meaningful identifiers, and use other
techniques to manage change.

In this paper, we ask whether it is possible to overcome these and
some related issues with monotonic, numeric identifiers while
remaining semantics-free. We describe our solutions, along with the
identitas library which implements these.

\textbf{Racing:} One unusual aspect of ontological
identifiers is that they are usually monotonically
increasing. This causes a significant race condition if two developers
are building a single ontology in parallel. If both attempt to add a
new term, they both must \emph{coin} a new identifier, which must be
unique. This is impossible to achieve without some degree of
co-ordination. One typical strategy is for developers have to
pre-coordinate to build the ontology by using pre-allocation
schema. For example, one developer allocated with the IDs from 1 to
1000, another allocated with 1000 to 2000 and so on. This approach is
effective, however it requires developers to manage the ID space
accurately, and also reduces the overall ID space since preallocated
IDs cannot be used elsewhere. Another approach is to just-in-time
co-ordinate; for example, the URIGen~\citep{urigen} server enables
this approach in Prote\'ge\'. Projects such as EFO (Experimental Factor Ontology)
and SWO (Software Ontology) use this to
manage their namespace. A final approach is to use temporary IDs, and
then allocate final IDs at a single, co-ordinated point in the
development process; URIGen also does this to enable off-line working.

We propose a much simpler approach which is to simply use random IDs
not just as temporary identifiers. While randomness does not \emph{a
  priori} completely remove the potential race condition, given a
large enough identifier space, the chances of collision can be reduced
to provide world (or universe) uniqueness. This approach is commonly
used with random UUIDs (Universal Unique Identifiers) being perhaps 
the most common example.

\textbf{Pronouncing:} The use of randomness raises a secondary
issue. These identifiers are likely to be relatively long,
exacerbatting the problems of memorability and pronounceability. One
solution to this problem is to just not show the identifiers to
humans. With tools like Prote\'ge\'\ this is possible, of course,
because it has a view which may be different from the underlying
model. With text file-formats, including OBO format, the various OWL
serialisations or the Tawny-OWL~\citep{lord_semantic_2013}
programmatic representation, this is rather harder (although the
latter does provide an mechanism for achieving this). It is also
difficult to do this for programmers developing tools like
Prote\'ge\', who are themselves using general tools such as IDEs,
debuggers and version control systems.

We have considered using a dictionary-based approach, to replace
numeric identifiers with English words. However, this approach raises
the probability of selecting a word which is inappropriate or
unfortunate -- consider the Sonic Hedgehog gene mutations which causes
holoprosencephaly in humans. Instead, we are investigating a solution
in the form of the
\emph{proquint}~\citep{DBLP:journals/corr/abs-0901-4016}. This is a
library build to encode numbers as a set of strings of alternating
consonants and vowels. Each consonant provide four bits of
information, each vowel only two bits, as shown in
Figure~\ref{fig:encoding}. Thus, sixteen bits can be represented using
five letters (3 consonants, 2 vowels).

\begin{figure}[htpb]
  \centering
  \begin{lstlisting}
     Four-bits as a consonant:
     0 1 2 3 4 5 6 7 8 9 A B C D E F
     b d f g h j k l m n p r s t v z

     Two-bits as a vowel:
     0 1 2 3
     a i o u
  \end{lstlisting}
\caption{Encoding bits as a proquint.}
\label{fig:encoding}
\end{figure}

For example a numeric identifier 10 associated with some term in a
given ontology would be translated to |babab-babap|, 11 would
be translated to |babab-babar| by using proquint function which
is quite readable, spellable and pronounceable string. In practice, if
used to represent random numbers, the proquints would rarely be so
close in alphabetic space. Note that proquints map directly to a
single number, so can be freely converted in either direction, and
that they are alphabetically ordered. Mappings between integer values
are shown in Figure~\ref{fig:int-to-proint}.

\begin{figure}[htpb]
\begin{center}
 \begin{tabular}{|c c|} 
 \hline
 Integer & Equivalent string \\ 
 \hline\hline
 0 & "babab-babab" \\ 
 \hline
 1 & "babab-babad" \\
 \hline
 2 & "babab-babaf"  \\
 \hline
 3 & "babab-babag"   \\
 \hline
 4 & "babab-babah"   \\
 \hline
 5 & "babab-babaj"    \\
 \hline
 Integer/MIN\_VALUE & "mabab-babab" \\
 \hline
 Integer/MAX\_VALUE & "luzuz-zuzuz" \\ 
 \hline
\end{tabular}
\end{center}
\caption{Integer to Proint(string).}
\label{fig:int-to-proint}
\end{figure}

In a simple extension, to the original algorithm, we have also
provided conversions from the Java short and long data types which
provides either a larger identifier space, or less typing; conversions
are shown in Figure~\ref{fig:short-long-pro}.

\begin{figure}[htpb]
\begin{center}
\small
 \begin{tabular}{|c c|} 
   \hline
   Short - Long & Equivalent string \\ 
   \hline\hline
   0 & "babab"  \\ 
   \hline
   1 & "babad"  \\ 
   \hline
   2 & "babaf"  \\ 
   \hline
   0 & "babab-babab-babab-babab" \\
   \hline
   1 & "babab-babab-babab-babad" \\
   \hline
   Long/MIN\_VALUE & "mabab-babab-babab-babab \\
   \hline
   Long/MAX\_VALUE & "luzuz-zuzuz-zuzuz-zuzuz \\
   \hline
 \end{tabular}
\end{center} 
\caption{Short and Long Convertion.}
\label{fig:short-long-pro}
\end{figure}

We note that the short range at $2^{16}$ numbers is large enough for
most ontologies current in operation. However, it is far too small
when combined with randomness as due to the birthday problem is very
likely to result in collisions even for small
ontologies~\citep{greycite81786}. The long range, meanwhile at
$2^{64}$ numbers is likely to cope for all ontological applications
where the identifiers are allocated as a result of human action; it
has half the bit-length of a UUID (which has a $2^{128}$ range).

\textbf{Checking:} We note that monotonic numeric ideas suffer from a
final problem. As well as being unmnenomic, if a numeric ID is
misunderstood, it is very likely that the incorrect ID is stil
actually a valid one; for instance, OBI:0001440 (``all pairs design'')
and OBI:0001404 (``genetic characteristics information'') are IDs
which differ in one one number.

A solution to this problem is well-understood with the use of a
checksum. For the identitas library, we use the Damm
algorithm~\citep{Michael_Damm_2007}. This algorithm is design to
operate on numbers, but it will work on proquints also, as they can be
converted to numbers. Examples of valid or invalid numbers are shown
in Figure~\ref{fig:valid}.
\begin{figure}
\centering
\begin{center}
\small
\begin{tabular}{|c c|} 
  \hline
  Random number & Validation True or False \\ 
  \hline\hline
  5724 & valid  \\
  \hline
  231  & invalid \\ 
  \hline
  0    & valid \\
  \hline
  222  & invalid \\
  \hline
\end{tabular}
\end{center} 
\caption{Validating generated IDs}
\label{fig:valid}
\end{figure}

Of course, the Damm algorithm incorporates a checksum so reduces the
total space of valid identifiers, in this case by an order of
magnitude, which will have implications if combined with
randomness. Under these circumstances, the larger numeric spaces (int
or long) are likely to be necessary.

In this paper we present a critique of current ontology semantics-free
identifiers; monotonically increasing numbers have a number of
significant usability flaws which make them unsuitable as a default
option, and we present a series of alternatives. We have provide an
implementation of these alternatives which can be freely combined. We
are now starting to integrate these into ontology development
environments such as Tawny-OWL~\citep{lord_semantic_2013}, and will
later provide an implementation for Prote\'ge\'. This form of
identifier space could significantly improve the management of
ontologies with very little cost.

\bibliography{phil_lord_refs}
\bibliographystyle{plain}

\end{document}